\documentstyle[prcrc,11pt,fleqn,psfig]{article}

\title{The impact on cosmology of a primordial scaling field}

\author{Pedro G. Ferreira\address{Center for Particle Astrophysics,
	301 Leconte Hall, \\ University of California, Berkeley, CA 94720 }%
        \thanks{This work was done in collaboration with Michael
	Joyce. Support came from NSF Cooperative
Agreement No. AST 9120005, PRAXIS XXI (Portugal) and
CNRS (France)}}

\begin{document}
\maketitle

\begin{abstract}
A scalar field with an exponential potential has 
the particular property that it is attracted into  
a solution in which its energy scales as the dominant component
(radiation or matter) of the Universe, contributing a fixed 
fraction of the total energy density.
We briefly discuss the dynamics of such a scalar field and its
impact on Big Bang nucleosynthesis, the growth of large
scale structure and abundance of damped Lyman$-\alpha$ systems
at high redshift. Given the simplicity of the model,
its theoretical motivation, and its success in matching observations,
we argue that it should be taken on par with other currently
viable models of structure formation.
\end{abstract}
\hspace{\fill}\\


The increasing quality of data from large scale surveys, cosmic
microwave background experiments, high redshift observations, 
and other cosmological measures, has begun to seriously
constrain the range of allowed models of structure formation.
In particular the simplest models have been shown incapable of
predicting the right level of fluctuations consistently on
GPc and MPc scales. The alternatives invoke
some form of fine tuning. This can be in the form of very
special initial conditions so that the energy density in a
cosmological constant, curvature, or some other exotic form of matter
comes to dominate at late times, or it can be in the form of
a non-clustering component that is motivated by standard particle
physics but whose mass scale is difficult to reconcile with 
the known particle bestiary.

Given the the unsatisfactory collection of models one has to
work with, it makes sense to find alternatives that might point
us in the right direction for finding a compelling and observationally
viable model of structure formation. It turns out that a general
class of theories in particles physics (Kaluza Klein, Supergravity,
or Superstring based) predict the existence of a weakly coupled scalar
field, $\Phi$, with a potential of the form

\begin{equation}
V(\Phi)=M^4_P\exp\left(-\lambda \frac{\Phi}{M_P}\right), \ \ \ \lambda \sim 
{\cal O}(1)
\end{equation}
where $M_P$ is the Planck mass, which has an attractor solution
(and therefore is insensitive to initial conditions) and leads to
a consistent level of fluctuations on all scales \cite{FJ97}. 

As shown in \cite{RP88,CLW93,W95} a scalar field with an exponential
potential (and $\lambda > 2$) will
evolve in such a way that its fractional energy density, $\Omega_\Phi$,
 is constant.
In particular one can show that
\begin{equation}
\Omega_\Phi=\frac{n}{\lambda^2}
\end{equation}
where  $n$ is 3 (4) in the matter (radiation) era. The competition between
the friction due to the expansion of the universe and the inertia
from the field rolling down the potential make it an attractor: if 
the energy density in $\Phi$, $\rho_\Phi$,
 is small compared to the total energy
density, it will remain constant until it becomes relevant. On the
other hand if $\rho_\Phi$ dominates, then it will evolve as
$\rho_\Phi\propto a^{-\lambda^2}$ until it catches up with the
background energy density. In the spirit of simplicity
we shall assume that the scalar field has been in the attractor
from very early times. In particular it contributed a non-negligable
fraction to the relativistic background at the time of nucleosynthesis.
This supplies us with a constraint on $\Omega_\Phi$ today:
\begin{eqnarray}
\Omega_\Phi<0.15-0.2
\end{eqnarray}

\begin{figure}[htb]
\begin{minipage}[t]{75mm}
\centerline{\psfig{file=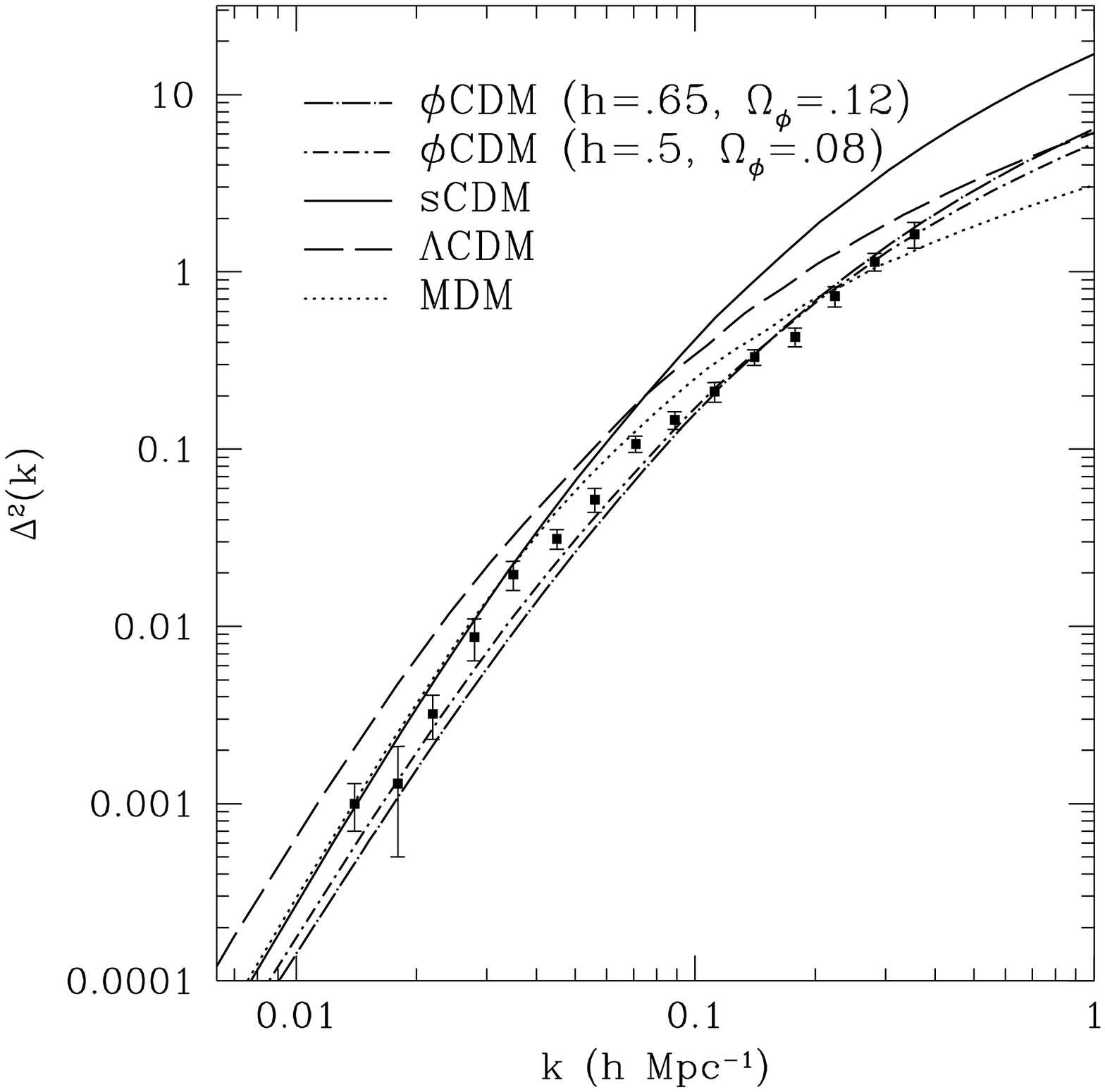,width=3.in}}
\caption{Mass variance per unit $\ln k$ computed from the Boltzman
code for different models compared with that inferred from a compilation of
galaxy surveys.}
\label{fig1}
\end{minipage}
\hspace{\fill}
\begin{minipage}[t]{75mm}
\centerline{\psfig{file=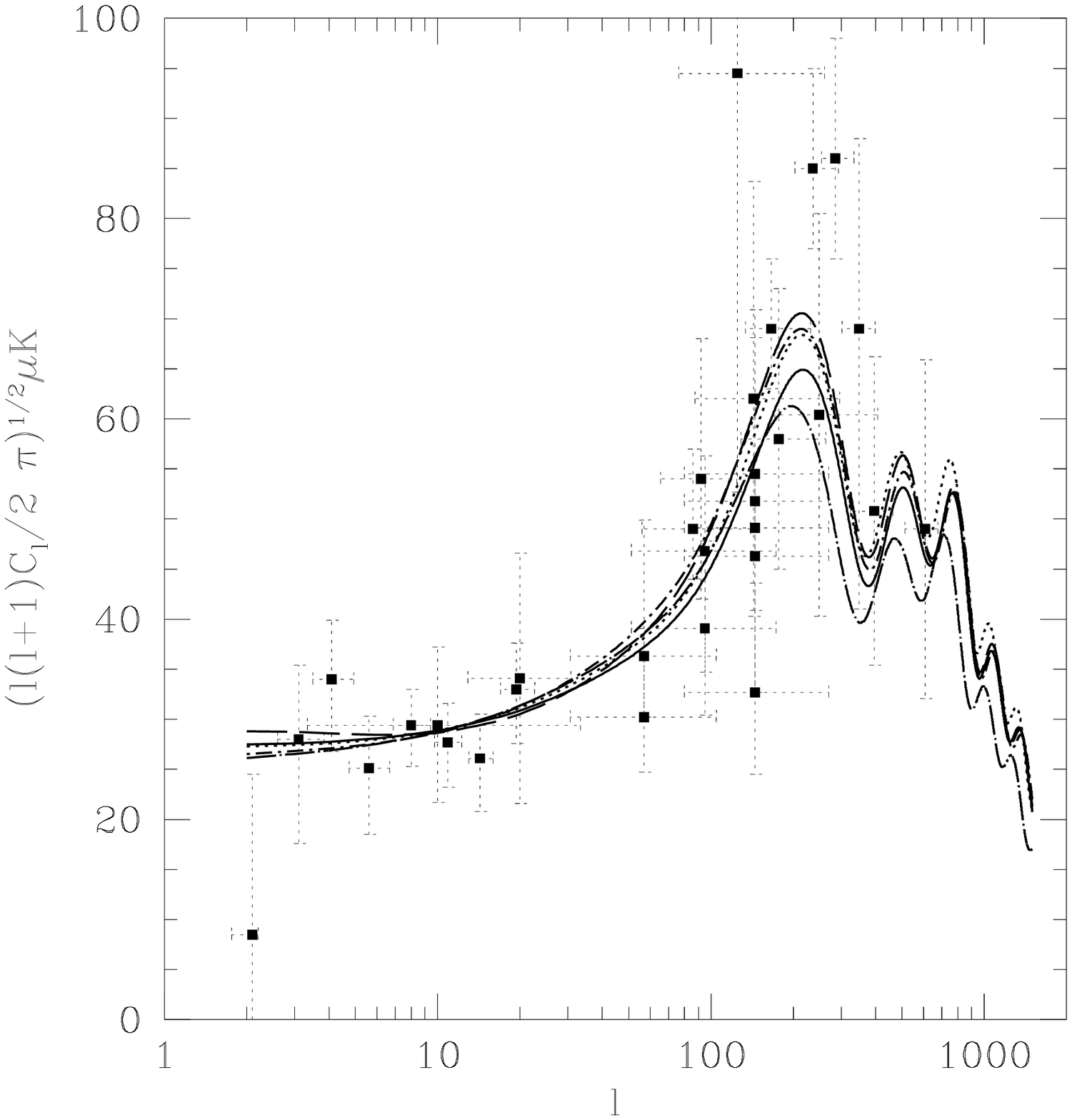,width=3.in}}
\caption{Comparison of different model predictions to current
experimental data. All models were COBE normalized and are labeled
as in Figure 1.}
\label{fig2}
\end{minipage}
\end{figure}

The fact that $\Omega_\Phi$ is so small might lead one to think that
such a scalar field has little or no impact on structure formation.
However the fact that such matter makes up a constant fraction of
the total energy density during the whole matter-dominated era make
its effects considerable. Indeed one can show that for an $H_0=50
(65)$ km
s$^{-1}$ Mpc$^{-1}$, an $\Omega_\Phi=0.08 (0.12)$ suffices to reconcile
the  cosmic microwave background data on horizon size scales with the
large scale structure data (see Figures \ref{fig1} and \ref{fig2}). We shall call such
a class of models $\Phi$CDM.

The evolution of perturbations in the presence of
the scalar field is simple to understand.
On superhorizon scales there is the usual growing mode with
$\delta_c \propto \tau^2$ (where $\delta_c$ is the 
density contrast in the CDM) and $\tau$ is conformal time.
 This is to be expected; the 
superhorizon evolution is insensitive to the ``chemistry'' of 
the matter and totally dominated by gravity. On sub-horizon scales in the
radiation era, the Meszaros effect comes into play giving
$\delta_c\propto \ln \tau$. The specific effect of the scalar
field appears on subhorizon scales in the matter era. 
The perturbation in the scalar field itself has the approximate 
solution $\delta_\Phi 
\propto (1/\tau^{3/2})J_{3/2}(k\tau)$ (where
$J_\nu$ is a Bessel function) which, when fed back into the equation
for $\delta_c$, gives an altered solution for the usual
growing mode $\delta_c \propto \tau^{2-\epsilon}$ where
\begin{eqnarray}
\epsilon=\frac{5}{2}\left(1-
\sqrt{1-\frac{24}{25}\Omega_{\Phi}}\right) \label{supp}
\end{eqnarray}
This solution shows explicitly how even a small contribution 
from the scalar field can give a significant effect, as it
acts all the way through the matter era.
The expected suppression of $|\delta_c|^2$  for  modes
larger than $k_{eq}$ is of order $(1+z_{eq})^{-\epsilon}$, where $k_{eq}$
is the wavenumber of the horizon size at radiation-matter equality. 
This last effect is reminiscent of the evolution of perturbations
in a mixed dark matter (MDM) universe where one has component
of matter, $\rho_\nu$, which is collisionless for a period of time during the
matter era.

There is, however a crucial difference between our model and MDM:
 the period of time during which
perturbations are suppressed is shorter in MDM compared
to $\Phi$CDM. In both cases there is a wavenumber, $k_{su}$,
which separates growing modes from damped modes. For $\Phi$CDM, 
this scale is roughly the horizon, i.e., $k_{su}\propto 1/\tau$, 
while for MDM it is the free streaming scale, i.e., 
$k_{su}=8a^{1/2}(m_\nu/10\ {\rm eV})h\ $Mpc$^{-1}\propto \tau$. 
Clearly in the latter case any given mode of $\delta_c$ will 
eventually start to grow. In particular, modes around $k_{eq}$
will already have started to undergo collapse. 

Finally, we find $\Phi$CDM has an interesting advantage 
over MDM. MDM models predict too little
structure  compared to that inferred from the Lyman-$\alpha$
measurements. From
Figure \ref{fig1} we can see that $\Phi$CDM should fare better than
MDM on very small scales. This is easy to understand: We argued 
 that effectiveness of $\Phi$CDM was mainly due to the fact
that the scalar field free-streaming scale grows with time, while
the massive neutrino free-streaming scale decays with time. 
We then need a larger amount of massive neutrinos to fit
both COBE and the cluster abundances in the MDM model than
the amount of scalar field in $\Phi$CDM. On much smaller scales
(the scales probed by Lyman-$\alpha$ systems), i.e., scales smaller
than the massive neutrino free-streaming scale, perturbations
in MDM should be more supressed than in $\Phi$CDM. This means
$\Phi$CDM should fare better than MDM with regards to the
Lyman-$\alpha$ constraints.

\end{document}